\documentstyle[11pt]{article}
\def\a{\alpha}
\def\b{\beta}
\def\g{\gamma}
\def\d{\delta}

\def\f{\phi}
\def\F{\Phi}

\def\y{\eta}

\def\m{\mu}
\def\n{\nu}
\def\r{\rho}
\def\p{\pi}

\def\t{\tau}
\def\v{\varphi}

\def\D{\Delta}

\def\para{\parallel}

\def\be{\begin{equation}}
\def\ee{\end{equation}}
\def\ol{\overline}
\def\ra{\rangle}

\begin{document}
\begin{titlepage}
\noindent
G\"{o}teborg ITP 95-21\\
September 1995\\
Revised: January 1996\\
hep-th/9509111\\

\vspace*{5 mm}
\vspace*{25mm}
\begin{center}{\LARGE\bf Gauge fixing, co-BRST and time evolution in QED}
\end{center}
\vspace*{3 mm} \begin{center} \vspace*{3 mm}
\begin{center}G\'eza F\"ul\"op\footnote{E-mail: geza@fy.chalmers.se}\\ 
\vspace*{7 mm} {\sl Institute of
Theoretical Physics\\ Chalmers University of Technology\\
G\"{o}teborg University\\
S-412 96  G\"{o}teborg, Sweden}\end{center}
\vspace*{25 mm}

\begin{abstract}
We consider BRST-invariant inner product
states for quantum electrodynamics
constructed from trivial BRST-invariant
states and a gauge regulator.
The trivial states are products of 
matter and ghost states and are annihilated by
hermitian operators.
The co-BRST
operator and some further gauge-fixing
regulators are found. The relation between  gauge fixing and 
time evolution of both the trivial and the 
inner product states is discussed.

\end{abstract} \end{center}
\end{titlepage}  
\setcounter{page}{1}
\setcounter{equation}{0}

\section{Introduction}

As discussed in \cite{batrob} for models with finite number of
degrees of freedom the physical inner product states can in general be obtained
by means of a gauge fixing regulator from  two simple solutions of the 
fundamental BRST-equation:
\be
Q \mid\v\rangle =0            \label{q}
\ee
where Q is the nilpotent hermitian BRST charge. 
These simple solutions 
 to equation (\ref{q}) are products of matter and ghost states: 
 $\mid \v_l \rangle \equiv \mid matter \ra \mid ghost \ra,  l=1,2$. 
In models with finite number
of degrees of freedom the  $\mid \v_l \rangle $ states
 were shown not to be well defined inner product states
by themselves in \cite{batrob}. 
They 
could be used however to obtain well defined inner product 
BRST-singlet states (denoted by $\mid s_l \rangle$)
by acting on them with some gauge-fixing regulators:
\be
\mid s_l \rangle = e^{\g_l K_l} \mid \v_l \rangle, \hskip 5mm l=1,2  
\label{1.2}
\ee
where $K_l = [\r_l, Q]$, $\r_l$ being a fermionic hermitian 
gauge-fixing operator
 and $\g_l$ is a real coefficient \cite{rob412}.
In \cite{roben} it was shown that the same singlets
can be obtained using linear combinations of the previous
operators $\a_l K_1 + \b_l K_2$ if a certian relation between
$\a_l, \b_l$ and $\g_l$ was satisfied. With some abuse of the
language we sometimes call the operators $K_l$ also
gauge-fixing operators, but this will never  lead to confusions
about which operators we mean. For non-abelian models the 
singlets $\mid s_l \rangle$ can no longer be written as products of purely
matter and ghost states.

The singlets are annihilated by  a set of  non-hermitian
operators. By the hermitian conjugates of these operators one 
can build a basis for the unphysical inner product states.
One can form  linear combinations of these inner product states
to get an orthogonal basis on the inner product state space.
The elements of this orthogonal basis are denoted by $\mid \xi \rangle$. 
Half of the $\mid \xi \rangle$ states have positive norm
and half of them negative. 
The hermitian metric operator ($\hat\y$) which
gives
\be
\langle \xi \mid  \hat\y  \mid \xi \rangle > 0, 
\hskip 1cm \forall \mid \xi \rangle
\label{eta}
\ee
defines the co-BRST operator by
\be
{}^{\ast} Q = {\y}^{-1} Q \y         \label{kob}
\ee
The physical space is completely determined by the BRST 
and the co-BRST operators
\cite{kalau}. One can always chose the metric operator to be
idempotent ($\y^2 =1$) and then the co-BRST operator becomes
${}^*Q = \y Q \y$.
Physical states are those that are annihilated by both these
operators.

One of the purposes with this paper is to find 
the appropriate generalizations
of the above results to the case of infinitely 
many degrees fo freedom and because
of this it is natural to consider first free electrodynamics.
The hope is to learn facts about the regulation procedure 
and the co-BRST also valid for more general field theories. 
The two simple BRST-invariant states (denoted also here
by $\mid \v_l \rangle$) are also here shown not to be inner
product states. By (\ref{1.2}) one can find the corresponding
inner product states ($\mid s \rangle$). As it will be shown later on,
every inner product state can  in this case be obtained from
both the $\mid \v_l \rangle$ states since
\be
\exp{(\g K_1)} \mid \v_1 \rangle = \exp{(\frac{1}{\g} K_2)} \mid \v_2 \rangle 
\ee In this sense the two trivial states $\mid \v_1 \rangle$ and
$\mid \v_2 \rangle$ are completely
equivalent. In non-Abelian theories the corresponding states have to be
treated somewhat differently. (See \cite{roben} for a model
with finite number of degrees of freedom.)
Using the method mentioned before (equations (\ref{eta}
and (\ref{kob})) 
one can find the co-BRST operator, in fact a whole class of them:
one for each value of $\g$. 
The $\mid s \rangle$ state found in equation (\ref{1.2}) is 
the physical vacuum. 
All the other
physical states can be built on it by physical 
creation operators. From the freedom in the choice
of the co-BRST charge one realises that there is a one-parameter
freedom in the choice of the vacuum state.

Since in electrodynamics we have an explicit non-vanishing
Hamiltonian one is also interested in the time-evolution
of both the trivial BRST-invariant $\mid \v_l \rangle$ and the inner
product $\mid s \rangle$ states.
It seems that the gauge fixing regulators are related to
time evolution. The Hamiltonian of free electrodynamics 
can be written as the sum of a BRST-closed and a BRST-exact term.
A formal similarity between the time evolution
operator on the equation of motion level and the
gauge-fixing regulator leads us to consider an operator
of the form $ \exp(H_{ph} + [Q, \r])$ and decompose it.
We are led to the conclusion that the vacuum singlets obtained before
are evolved in time as expected by a physical Hamiltonian
which only depends on the orthogonal components of the electric
field and the magnetic field. 
The non-physical part of the Hamiltonian is 
a simple combination of the gauge-fixing operators $K_l$.
The time-evolution operator built on this Hamiltonian
in an imaginary-time formalism is a generalized gauge-regulator.
The trivial $\mid \v_l \rangle$ states are exactly defined 
by the operators (four for each state) annihilating
them. One of these operators for each trivial state
is a gauge-fixing operator, sometimes called gauge-slicing.
As it is shown in Section 4 there exists a gauge-slicing
such that the state it defines evolves under an
imaginary-time evolution operator equivalent to the
gauge regulator itself.

There exist several features that differ in this paper
as compared to previous works. A fundamental  analysis
of quantum electrodynamics and Yang-Mills theories
was given by Kugo and Ojima in \cite{kugo}.
The main difference between \cite{kugo} and the
present paper is the fact that in \cite{kugo} one
uses a perturbative Fock-space approach.
One builds all states on a vacuum that is itself an inner
product state.
The present paper uses a   non-perturbative approach, 
the fundamental states being eigen-states of the
connections (scalar and vector potentials) and the electric
fields.
We look for BRST-invariant states of the form
$\mid matter \ra \mid ghost \ra$.
This leads us as proven in \cite{batrob}, \cite{rob412}
to states (including the vacuum state) that are not
well defined inner product states.
Since this situation occurs quite generally it can be instructive
to first analyze a simple case like quantum electrodynamics.
In the physical state space as defined in \cite{kugo}
it is the quartet mechanism that guarantees that unphysical
states always appear in zero-norm combinations, thus
not affecting any physical inner products.
In the present paper we define the physical inner
product states as the one being annihilated by both
the BRST and the co-BRST operators.
This definition does not leave room for any non-physical
states in the physical state space and not even zero-norm 
states are left.

In Section 2 we first   introduce
the formalism which is built on the eigen-vectors of the
connection (the vector and scalar potential). 
This basis is very convenient since every operator
in the Hamiltonian formulation corresponds to either the
connections or to their momenta. For more complicated models
 the non-perturbative nature of these eigen-vectors can
also be an advantage.
We explicitly prove that the trivial BRST-invariant
states, that is solutions to eq.(\ref{q}) denoted by $\mid \v_1 \rangle$
and $\mid \v_2 \rangle$, for free electrodynamics are not inner
product states.
More exactly it is shown that $\langle \v_l \mid \v_l \rangle = 0 \infty,
\hskip 5mm l=1,2$.

In Section 3 we find the well defined inner product states
using the prescription given by Batalin and Marnelius in \cite{batrob}.
We show that these states are also normed to unity if the eigen-value of the
time-like component of the connection (the scalar potential)
operator is imaginary.
It is shown that the BRST and co-BRST operators together
completely define the singlet vacuum  state, as expected from \cite{kalau}.

The two gauge fixing operators $K_l, l=1,2$ are members of 
an $SL(2,R)$ algebra,
thus one can use (whenever convenient) linear combinations of
them instead of any of them. These details are discussed in
Section 4 and the conclusions we arrive at
are essentially the same as in the case of  finite number
of degrees of freedom \cite{roben}.

The second part of Section 4 deals with the time evolution of the
$\mid s \rangle$ and the $\mid \v_l \rangle$ states.
It contains the decomposition of the Hamiltonian
into a physical part and a gauge-fixing term.
We show a way how one can understand this decomposition
in an imaginary-time formalism.
We also comment on a possible interpretation
in the real time case related to the path-integral
formalism.
The Appendix contains some
details of the decomposition of the unphysical Hamiltonian
into a combination of the gauge-fixing operators $K_l$
and another unphysical operator.

\setcounter{equation}{0}
\section{The BRST-invariant states}                                

The formalism used throughout this paper is built on
the eigenvectors of the connection ($A^{\m}(x)$) and the
fermionic ghost and anti-ghost  ($\y(x)$ and $\bar{\y}(x)$) operators:
\be
\mid A \rangle = \prod_{\vec{x}} \prod_{\m} \mid A^{\m}(\vec{x}) \rangle, 
\mid \y \rangle = \prod_{\vec{x}} \mid \y(\vec{x}) \rangle,
\mid {\ol\y} \rangle = \prod_{\vec{x}} \mid {\ol\y}(\vec{x}) \rangle
\label{2.1}
\ee
where
\be
\hat A^{i}(x) \mid A^i(\vec{x})\rangle  = 
A^{i}(x) \mid A^i(\vec{x})\rangle  \label{ai}
\ee
\be
\hat A^{0}(x) \mid A^0(\vec{x})\rangle  = 
i A^{0}(x) \mid A^0(\vec{x})\rangle \label{a0}
\ee
\be
\hat{\y}(x) \mid \y(\vec{x}) \rangle = 
{\y}(x) \mid \y(\vec{x}) \rangle    \label{gh}
\ee
\be
\hat{\ol{\y}}(x)\mid\y(\vec{x})\rangle =
{\ol{\y}}(x) \mid \y(\vec{x}) \rangle\label{agh}
\ee
There are two reasons for this choice. One of them is that
the connections, the ghosts and their momenta 
are the fundamental variables of the
theory both in the Lagrangean and the Hamiltonian
formalism. In the Hamiltonian formalism all the operators,
including the BRST-operator, are expressed as functionals
of the operators that correspond to the fundamental variables.
The other reason is the non-perturbative nature of such eigenstates
in a more general context.
As there exist several areas (like quantum gravity
(see e.g. \cite{ashtekar})) where perturbative methods do not lead
to renormalizable theories one is lead to study some
intrinsically non-perturbative formalisms
(\cite{kiefer}).

The eigenvalue of the time component of the
connection is imaginary (\ref{a0}) because this is necessary
for the 
normalizability of the states.
Vectors of this kind were first introduced by Pauli \cite{pauli}.
For a detailed analysis see e.g.\cite{arisue}. The connection eigen-states
(\ref{ai} and \ref{a0}) form a complete and orthonormal basis:
\be
\langle A' \mid A \rangle = \prod_{x} {\d}^3\left({A'}^i(x) - {A}^i(x)\right) 
\d \left({A'}^0(x) + {A}^0(x)\right)   \label{a'a}
\ee
The last term follows from the imaginary eigenvalue of $A^0$:
\be 
\langle A' \mid \hat{A}^0 \mid A \rangle = 
-i {A'}^0(x)\langle A' \mid A \rangle 
= \langle A' \mid A \rangle i {A}^0(x)     \label{a1}
\ee

The eigenvalue equations of the ghost operators are defined in the same way
as the equations of the matter operators
(in equations (\ref{gh}) and (\ref{agh})) 
but there exist certain differences
due to the fermionic nature of the ghosts.
 For some details about  eigenvalue equations of fermionic
operators see e.g.\cite{robferm}. One of their important properties
that we have to keep in mind is the fact that the $\d$-function of
a fermionic variable is proportional to the variable itself:
\be
\d(\y - \y') =  (\y - \y')
\ee
This relation results in the following remarkable property of the
ghost inner product states:
\be
\langle \y \mid \y' \rangle \propto \d(\y - \y') = -i (\y - \y')   \label{yy}
\label{ghinner}
\ee which vanishes for $\mid\y\rangle = \mid\y'\rangle$.
A general wave functional expressed in the  basis (\ref{2.1}) is the
Schr\"odinger representation:
\be
\F[A,\y,\overline{\y}] = \langle A, \y, \ol{\y} \mid\v\rangle
\ee
(Throughout this paper both the wave-functional and the ket 
forms are going
to be used depending on which of them is more convenient.)

The physical states in a theory are expressed by the 
cohomology classes of the BRST-operator i.e. classes of 
states that are eliminated
by it but which cannot be written as the BRST-operator acting on
any other state. When solving the equation $Q\mid\f\rangle = 0$
one finds two different classes of solutions built on different
ghost vacua. These states are not inner product states
as it was proved in
\cite{batrob} for systems with finite number of degrees of
freedom. As we shall see in this section the situation
is the same in the case of electrodynamics too.

An important assumption in \cite{hw-marn} 
was that one can always write the wave functionals 
corresponding to physical states
as products of a matter and a ghost part:
\be
\F[A,\y,\overline{\y}] = \langle A\mid \v\rangle_{m} 
\langle\y,\overline{\y}\mid \v\rangle_{gh} \equiv
\F_m[A] \F_{gh}[\y, \ol{\y}] \label{decomp}
\ee
This decomposition, together with the completeness of the
$\mid A \rangle$ basis leads us to the formal
inner product for the matter part
\be
\langle\v \mid \Psi\rangle = \int {\cal D}A {\F}^{\ast}_m[A] {\Psi}_m[A] 
\ee
where ${\cal D}A = \prod_{\vec{x}} \prod_{\m} dA^{\m}(\vec{x})$.
The ghost part does not yield a well defined inner product
because of the properties described in eq.(\ref{yy}).
Another natural and much more geometrical way to describe 
BRST-quantization would be to work on the space of all the 
connections. The BRST-operator acts in
this space as an exterior derivation operator. (See e.g. 
\cite{nakahara}, \cite{henneaux} and for an example \cite{witten}.)

The Lagrangian we are going to use is:
\be
{\cal L}= -\frac14 F_{\m\n}F^{\m\n} -\frac12({\partial}_{\m} A^{\m})^2
-i {\partial}_{\m} \overline{\y}  {\partial}^{\m} \y        \label{L}
\ee                               
The imaginary unit ($i$) appears because we demand that both the Lagrangian
and the ghost variables 
are real. One could of course redefine
the ghosts to be complex conjugated to the anti-ghosts and then the $i$ would
disappear. The reality of the ghost variables  causes their momenta to be
purely imaginary. In the Hamilton formalism the non-zero commutators
between the fundamental operators are as follows:
\be
[A^{\m}(x), E^{\n}(y)]_{x^0 = y^0} = i g^{\m\n} \d^3(\vec{x}-\vec{y} )
\ee
\be
[\y(x), {\cal P}(y)]_{x^0 = y^0} = i \d^3(\vec{x}-\vec{y} )
\ee
\be
[\ol{\y}(x), \ol{ {\cal P}}(y)]_{x^0 = y^0} = i \d^3(\vec{x}-\vec{y} )
\ee
where $g^{\m\n}$ is the Minkowski metric Diag$[-1,1,1,1]$.
The BRST charge of free electrodynamics in the phase space
\cite{batvilk} is:
\be
Q = \int d^3x \left(\partial_i E^i(x) \y(x) 
-i E^0(x) \ol{\cal P}(x)\right)    \label{Q}
\ee
This somewhat unconventional form of the term containing the 
anti-ghost momenta multiplied by the imaginary unit insures the
hermiticity of the BRST-charge.
This BRST-charge is only nilpotent using the equations of 
motion. One also notices that this Lagrangian 
is anti-BRST invariant too, the
anti-BRST charge in abelian models being of the same form as 
the BRST-charge with the
ghost and anti-ghost variables (and momenta) interchanged:
\be
\ol{Q} = \int d^3x [\partial_i E^i(x) \ol{\y}(x) 
+i E^0(x) {\cal P}(x)]                                         \label{antiQ}
\ee
For more details about the anti-BRST charge and its role
in quantizing gauge theories
see e.g.\cite{curci}.
As a first step in searching for the physical states
we are looking for the BRST-invariant and anti-BRST invariant 
ones, i.e. those satisfying:
\be
Q \F[A,\y, \ol{\y}] = 0      \label{qonf}
\ee                                                                 
\be
\ol{Q} \F[A,\y, \ol{\y}] = 0      \label{antiqonf}
\ee
Using the decomposition of the wave-functional into
a matter and a ghost part from eq.(\ref{decomp}) one obtains the
two classes of solutions mentioned in the introduction.
The {\bf first} class of solutions are those states that satisfy:
\be 
E^0(\vec{x}) \F_{1m}[A] =\y(\vec{x}) \F_{1gh}[\y, \ol{\y}] = 
\ol{\y}(\vec{x}) \F_{1gh}[\y, \ol{\y}] =0                        \label{elso}
\ee
The {\bf second} class is given by:
\be
\partial_i E^i(\vec{x}) \F_{2m}[A] = 
\ol{\cal P}(\vec{x}) \F_{2gh}[\y, \ol{\y}] =
{\cal P}(\vec{x}) \F_{2gh}[\y, \ol{\y}] = 0                   \label{gauss}
\ee
There are two more conditions, one for each state, that 
can be consistently imposed
on these states and they are gauge-fixing conditions of the form \cite{batrob}
\be
-\frac{\partial_i}{\D} A^i(x) \mid \v_1 \rangle =0              \label{elso2}
\ee
\be
A^0(x) \mid \v_2\rangle =0                                    \label{masodik2}
\ee
The matter part of the {\bf first} equation (\ref{elso}) 
only tells us that these states do
not depend on $A^0$. The ghost part of (\ref{elso}) 
gives the ghost vacuum as solution.
That is the eigenvalues of both the ghost and 
the anti-ghost operators
vanish all over the space. It follows then that the  states
corresponding to the solution of 
the eq.(\ref{elso}) are of the form
\be
\mid \v_1 \rangle = 
\mid \v[\vec{A}]\rangle \otimes  \int {\cal D}A^0 \mid A^0\rangle 
\otimes \mid 0\rangle_{\y \ol{\y}}
\ee
These states are obviously not well defined inner
product states. $\langle ~_1\v \mid \v_1\rangle$ results in $\infty 0$ in 
every space point, the infinity comes
from the integral by $dA^0$ and the zero from the ghost 
part (see eq.\ref{ghinner}).

The {\bf second}
 class of solutions to the BRST- and anti-BRST equations is given
by eq.(\ref{gauss}).
The ghost parts of the equation only tell us that the ghost part 
of the state 
does not depend on the ghost $\y$ and the 
anti-ghost $\ol{\y}$.  The ket corresponding to
this solution can then be written as
$\int {\cal D}{\y} \mid \y\rangle \otimes 
\int{\cal D}\ol{\y} \mid \ol{\y}\rangle $.
One can also understand these states as the ghost-momentum-vacuum.
The matter part of equation (\ref{gauss}) can easily be solved if 
one goes over to the conjugated $E$ representation:
\be
\F_m[A] = \int {\cal D}E \langle A \mid E\rangle \langle E \mid \v\rangle = 
\int  {\cal D}E e^{\frac{i}{\hbar} \int d^3x A_{\m}(x) E^{\m}(x)} \F[E]
\ee
If $\F_m[A]$ is to satisfy (\ref{gauss}) $\F[E]$ has to be of the form
$\prod_{x} \d({\partial}_i E^i(x)) \tilde{\F}[E]$. By writing the naive 
integral measure as: $dE^0 d^2E^{\perp} dE^{\parallel}$ and using the fact
that:
\begin{displaymath}
E^{\parallel} = \frac{\partial_i}{\sqrt{-\D}} E^i
\end{displaymath} one obtains:
\be
\F_m[A] = \int {\cal D}E^0 {\cal D}^2E^{\perp} \frac{1}{\sqrt{-\D}}
\tilde{\F}(E^0, E^{\perp}, E^{\parallel}=0) e^{\frac{i}{\hbar} 
\int d^3x (A_0 E^0 + A^{\perp} E^{\perp})}
\ee
The ${\sqrt{-\D}}$ operator is a well-defined real and positive operator
in the momentum space representation. Since
the space-time metric is $g_{\m \n} =$ Diag$(-1, 1,1,1)$ 
$(-\D)$ corresponds to $k_i k^i$ in the momentum space.   
The obvious problem with this solution is the same as in case 1: 
it is not a well defined inner
product state. One can see this by taking the naive inner product
between two states of the form $ \mid \F\rangle = 
\int {\cal D}A \F_m[A] \mid A\rangle
 \int d{\y} \mid \y \rangle \int d{\ol\y} \mid \ol\y \rangle $. 
The result is again an ambiguous
$\infty 0 $, where the infinity comes from the integral over the parallel
component of the connection and the zero comes from the 
(anti-)ghost part of the
integral following from eq.(\ref{yy}). 
The same result can be obtained even without introducing
the $\mid E\rangle$ kets. One can simply chose another base for the connection
and write the integral measure decomposed into a time-like, an orthogonal and
a parallel part. The matter part of the state becomes:
\be
\mid \F\rangle_m = 
\prod_x \int dA^0 d^2A^{\perp} dA^{\parallel} \F[A] \mid A^0\rangle 
\mid A^{\perp}\rangle
\mid A^{\parallel}\rangle
\ee
The constraint expressed in (\ref{gauss}) means essentially 
that $\F[A]$ does not depend on the parallel component of $A$: 
it has to be a function of e.g.
$A_i(g^{ij}-\frac{{\partial}^i {\partial}^j}{\D})A_j 
= A^{\perp} A^{\perp}$. For a more detailed description see
\cite{jackiw}. So neither of the two trivial solutions to the
BRST equation (\ref{q}) is an inner product state.

\setcounter{equation}{0}
\section{Inner Product States}

To summarize what has been  said up till now: the fundamental equation
($Q \mid \v \rangle =0$) has two trivial solutions, $\mid \v_1 \rangle$ and
$\mid \v_2 \rangle$ defined by the relations (\ref{elso}) and (\ref{elso2})
resp. (\ref{gauss}) and (\ref{masodik2}) or expressed in the ket-notation:
\be
\partial_i A^i(x) \mid \v_1\rangle = E^0(x)\mid \v_1\rangle = 
\y(x) \mid \v_1\rangle 
= \ol{\y}(x)\mid \v_1\rangle =0       \label{fi1}
\ee
\be
A^0(x)\mid \v_2\rangle = \partial_i E^i(x) \mid \v_2\rangle= 
{\cal{P}}(x)\mid \v_2\rangle 
= \ol{\cal{P}}(x)\mid \v_2\rangle = 0       \label{fi2}
\ee
None of the solutions of these conditions is an inner product state. 
Since physical
states have to be inner product states we have to find such states built
on the solutions we obtained. The inner product states have to be also
BRST-invariant.

The way this problem is dealt with in models with finite degrees
of freedom \cite{batrob} is that one acts with  gauge fixing regulators
on the two $\mid \v_i\rangle$ states. These gauge fixing regulators
are the exponentials of commutators between the BRST operator and
a gauge-fixing condition so the BRST invariance
of the new states is automatically guaranteed.

In the case of free electrodynamics
the inner product states will be of the form
\be
\mid s_l\rangle = e^{\g_l[\r_l,Q]}\mid \v_l\rangle,    
\hskip 1cm   l=1,2  \label{singl}
\ee
The gauge-fixing operators in this case are:
\be
\r_1 = -i \int d^3x \frac{1}{\sqrt{-\D}} A^0(x) {\cal P}(x)
\ee
\be
\r_2 = \int d^3x \frac{\partial_i}{\sqrt{-\D}} A^i(x) \ol{\y}(x) 
=  \int d^3x A^{\parallel}(x) \ol{\y}(x)
\ee
and $\g_l$ are just constants. (There is no summation 
over the indices $l$.) The operators involving $\sqrt{-\D}$
are defined as in the previous section by their correspondents
in the momentum space. $A^{\parallel}(x)$ is the parallel
component of the gauge-field $A^{\m}(x)$. Its  conjugated
momentum is:
\be
E^{\parallel}(x) = \frac{\partial_i}{\sqrt{-\D}} E^i(x)
\ee
and the equal-time commutation relation between them is
\be
[ A^{\parallel}(x), E^{\parallel}(y)] = i \d^3(\vec{x}-\vec{y})
\ee  
>From here one easily obtains:
\be
K_1 \equiv [Q,\r_1] = \int d^3x \left( A^0(x) E^{\parallel}(x) 
+ i {\cal P}(x) \frac{1}{\sqrt{-\D}} \ol{\cal P}(x)\right)          \label{k1}
\ee
\be
K_2 \equiv [Q,\r_2] = \int d^3x \left(E^0(x) A^{\parallel}(x) 
+ i \ol{\y}(x) \sqrt{-\D} \y(x)  \right)                            \label{k2}
\ee
The operators that annihilate the singlet states obtained in
this way are given by
\be
{D'}_l = e^{\g_l K_l} D_l e^{-\g_l K_l}
\ee
where $D_l$ stands for all the operators annihilating
$\mid \v_l\rangle$. Again there is no summation over $l$  included.
Since from $\partial_i A^i \mid \v_1\rangle =0$ follows that
$A^{\parallel} \mid \v_1\rangle =0$ one can use $A^{\parallel}$ in
eq.(\ref{fi1}). The situation is of course the same 
for the $\partial_i E^i$ in the case of the second state: one can
replace it by $E^{\para}$.
The conditions which $\mid s_1\rangle$ satisfies are:
\begin{displaymath} 
\left(A^{\parallel}(x) - i \g_1 A^0(x)\right) \mid s_1\rangle = 
\left(E^0(x) - i \g_1 E^{\parallel}(x)\right)\mid s_1\rangle=
\end{displaymath}
\be
=\left(\y(x) +\g_1 \frac{1}{\sqrt{-\D}}\ol{\cal P}(x)\right) \mid s_1\rangle =
\left(\ol{\y}(x) - 
\g_1 \frac{1}{\sqrt{-\D}}{\cal P}(x)\right)\mid s_1\rangle =  0      \label{s1}
\ee
while the conditions on $\mid s_2\rangle$ are:
\begin{displaymath}
\left(A^0(x) + i \g_2 A^{\parallel}(x)\right) \mid s_2\rangle = 
\left(E^{\parallel}(x) + i \g_2 E^0(x)\right)\mid s_2\rangle =
\end{displaymath}
\be
=\left({\cal P}(x) - \g_2 {\sqrt{-\D}}\ol{\y}(x)\right) \mid s_2\rangle = 
\left(\ol{\cal P}(x) - \g_2 {\sqrt{-\D}}{\y}(x)\right) \mid s_2\rangle = 0     
\label{s2}
\ee
It is interesting to note that for the special choice of the
constants $\g_1 \g_2 = 1$ the two singlet states are identical.
This means that any singlet can be reached from
both trivial states and there exists a one-to-one map
between the two trivial solution spaces containing $\mid s_1\rangle$
and $\mid s_2\rangle$ .
It is enough therefore to denote a singlet only
by $\mid s\rangle$ and keep in mind that it can be obtained
from both  $\mid \v_i\rangle$-s by different $\g_i$-s.
The only time when the notation  $\mid s_i \rangle$ is useful
is when one wants to explicitly show how one obtained that state.

An important question is under what conditions are
the singlet states normalized. The normalization conditions
led to some explicit values  for the  $\g_i$-s.
To find the value of $\g_1$ one can rewrite the annihilation
operators in equations (\ref{s1})  in a somewhat
simpler form by introducing the following operators:
\be
\v(x) = - E^0(x) + i \g_1 E^{\para}(x)
\ee
\be
\psi(x) = A^{\para}(x) - i \g_1 A^0(x)
\ee
\be
\r(x) = i \sqrt{-\D}\y(x) + i \g_1 \ol{\cal P}(x)
\ee
\be
k(x) = - \ol{\y}(x) + \g_1 \frac{1}{\sqrt{-\D}}{\cal P}(x)
\ee
with the (anti)commutation relations:
\be
[\v(x), \psi(y)] = 0      \label{com1}
\ee
\be
[\v(x), {\psi}^{\dagger}(y)] = 2 \g \d^3(x-y) \label{com2}
\ee
\be
[\r(x), k(y)] = 0                          \label{com3}
\ee
\be 
[\r(x), k^{\dagger}(y)] = 2 \g \d^3(x-y)       \label{com4}
\ee
The BRST-charge in this formulation is
\be
Q= \frac{1}{2 \g} \int d^3x[\v(x) \r^{\dagger}(x) + \v^{\dagger}(x) \r(x)]
\label{brstfi}
\ee
while the singlet is going to be defined by:
\be
\v(x) \mid s_1\rangle = \psi(x) \mid s_1\rangle = \r(x) \mid s_1\rangle =
 k(x) \mid s_1\rangle =0
\ee
The simplest way to see whether a state given by
the equation (\ref{s1})  is normalized or not is
to use the wave-functional representation 
($\Psi[A,\y,\ol{\y}] \equiv \langle A, \y, \ol{\y} \mid s\rangle$). 
The first two equations in  (\ref{s1}) are in this formulation:
\be
\left(A^{\para}(x) - i \g A^0(x) \right) \Psi[A, \y, \ol{\y}] = 0       
\label{a}
\ee
In order to find a solution to the equation (\ref{a}) one has
to presume that the operators $\hat{A}^0(x)$ have imaginary eigenvalues
as mentioned in Section 2:
\be
\hat{A}^0(x) \mid i A^0 \rangle = i A^0(x) \mid i A^0 \rangle
\ee
where $A^0(x)$ is real.
The formal inner product based on (\ref{a'a}) gives 
the $A$-dependent functionals a norm:
\be
\int {\cal D}^4A \Psi^{\ast}[\vec{A}, -A^0] \Psi[\vec{A}, A^0]
\ee The solution
to equation (\ref{a}) is given by
\be
\Psi[A^{\para}, iA^0, \y, \ol{\y}] \propto 
\prod_x \d(A^{\para}(x) + \g_1 A^0(x))
\label{ps}
\ee
Because of the imaginary eigenvalue of $\hat{A}^0$ the operator 
equation corresponding to the second equation in (\ref{s1}) becomes:
\be
\left( \g_1 \frac{\d}{\d A^{\para}(x)} -  
\frac{\d}{\d A^0(x)} \right) \Psi[A, \y, \ol{\y}] = 0 
      \label{e}
\ee
We see that (\ref{ps}) is a solution to this equation too.
The last two equations in (\ref{s1}) result in another proportionality:
\be
\Psi[A^{\para}, iA^0, \y, \ol{\y}] \propto 
\prod_x \d(\sqrt{-\D} \y(x) + \g_1 \ol{\cal P}(x))
\ee
(Note that here the fact that
the ghost momenta are anti-hermitian eliminates the need for any
other imaginary eigenvalues.) 
Computing now the norm of the vacuum state one obtains
\be
\langle \Psi \mid \Psi \rangle = \prod_{\vec{x}} \frac{1}{\g_1} \g_1 = 1
\ee
In this simple case one can understand this result in the
following way: the wave-functional can be written as the product
of a matter and a ghost part. The norm of the matter part is 
$\prod_{\vec{x}} \frac{1}{2 \g_1}$ while the norm of the ghost state
is ${2 \g_1}$.
Starting from the other singlet one obtains the same result
as expected. What we have found now is that every inner product vacuum state
is normalized to unity. We also notice that in our case
the trivial solutions could be written as products of a matter
and a ghost state. Since the gauge regulator $\exp{(K_l)}$
can also be decomposed in a matter and a ghost factor that 
commute with each other even the inner product singlet is a 
product of a matter and a ghost functional. 

One can now find the orthogonal creation and annihilation
operators that will eventually lead to the explicit form of
the co-BRST operator. One needs to find those combinations of
$\v, \psi, \r, k$ that give a positive and a negative
definite matter (ghost) operator in every space point, that is
combinations that are subject to the following (anti)commutation
rules:
\be
[D(x), D^\dagger(y)] = - [F(x), F^\dagger(y)] = \d(x-y)
\ee
\be
[D(x), F^\dagger(y)] =0
\ee
\be
[G(x), G^\dagger(y)]_+ = - [H(x), H^\dagger(y)]_+ = \d(x-y)
\ee
\be
[G(x), H^\dagger(y)]_+ =0
\ee
where $D(x)$ and $F(x)$ are bosonic while $G(x)$ and $H(x)$ are
fermionic annihilation operators. The coefficients in these
linear combinations need not be constants, they can be coordinate
dependent. If
\be
D(x) = d_1(x) \v(x) + d_2(x) \psi(x)
\ee
\be
F(x) = f_1(x) \v(x) + f_2(x) \psi(x)
\ee
\be
G(x) = g_1(x) \r(x) + g_2(x) k(x)
\ee
\be
H(x) = h_1(x) \r(x) + h_2(x) k(x)
\ee
then the (anti)commutation relations given before lead to:
\be
\mid d_1(x) \mid = \mid f_1(x) \mid \hskip 1cm \mid d_2(x) \mid = 
\mid f_2(x) \mid 
=\frac{1}{4 \g \mid d_1(x) \mid}
\ee
\be
\mid g_1(x) \mid = \mid h_1(x) \mid \hskip 1cm \mid g_2(x) \mid = 
\mid h_2(x) \mid 
=\frac{1}{2 \mid h_1(x) \mid}
\ee
For notational simplicity from now on we do not explicitly denote
the coordinate dependence of the operators.
Restricting ourselves to real coefficients we obtain:
\be
D= d_1 \v + \frac{1}{4 \g d_1} \psi
\ee
\be
F = \pm (d_1 \v - \frac{1}{4 \g d_1} \psi)      \label{F}
\ee
\be
G(x) = g_1 \r + \frac{1}{4 \g g_1} k
\ee
\be
H = \pm (g_1\r - \frac{1}{4 \g g_1} k)           \label{H}
\ee
It is clear that $F^{\dagger}$ and $H^{\dagger}$ create the negative
normed states. There exists a class of metric operators that
define new inner products such that every state is 
positive definite \cite{kalau}:
\be
\langle \xi \mid \hat{\y} \mid \xi \rangle > 0,    \forall \mid \xi\rangle
\ee This metric operator is given by:
\be
\hat{\y} = e^{i \pi \int d^3x [F^{\dagger} F - H^{\dagger} H]}
\ee
The BRST operator in these variables for the positive sign solutions of
(\ref{F})  is
\be
Q= \int d^3x \frac{1}{8 \g d_1 g_1} [(D^{\dagger} + F^{\dagger})
(G+ H) + (D+ F)(G^{\dagger} + H^{\dagger})]
\ee
The overall factor can always be chosen to be unity because it 
would not change the physical effect of the operator at any space point.
The co-BRST is given by
\be
{}^{\ast}{Q} \equiv \hat{\y} Q \hat{\y} =
\int d^3x [(D^{\dagger} - F^{\dagger})
(G- H) + (D - F)(G^{\dagger} - H^{\dagger})]
\ee
or expressed in the variables used in the definition of the
singlets
\be
{}^{\ast}{Q} = \int d^3x \frac{1}{\g} [\psi k^{\dagger}
+\psi^{\dagger} k]
\ee
which is perfectly consistent with the definition of the
singlet states as being the ones eliminated by
$\v, \psi, k, \r$. The same result is obtained for the negative sign 
solutions in (\ref{F}) and (\ref{H}). A similar expression was also
found in \cite{yang}. Thus we come to the same conclusions
as \cite{kalau}, namely that the singlet states are entirely
defined by:
\be
Q \mid s\rangle = {}^{\ast}Q \mid s\rangle = 0
\ee
Returning now to the original variables the co-BRST operator
takes the form
\be
{}^{\ast}Q = \int d^3x [-\frac{1}{\g} A^{\para} \ol{\y} + 
i \g A^0 \frac{1}{\sqrt{-\D}} {\cal P}]
\ee
We notice that for every choice of the coefficient $\g$
there exists a co-BRST operator. In other words by fixing the co-BRST
operator, that is by chosing $\g$ we uniquely define the vacuum
singlet state.

\setcounter{equation}{0}
\section{Generalized Gauge Fixing and Time Evolution}

In the previous section we saw how the operators $\exp{(\g_i K_i)}$ 
acting on the trivial BRST-invariant $\mid \v_i\rangle$ states
resulted in well defined inner product states denoted first as
$\mid s_i \rangle$. We noticed then that every singlet state $\mid s_i \rangle$
could be reached from both original $\mid \v_i\rangle$-s:
\be
\mid s_i \rangle = \exp{(\g_1 K_1)} \mid \v_1\rangle = 
\exp{(\frac{1}{\g_1} K_2)} \mid \v_2\rangle
\ee
Since $\g \neq 0$ and $\mid \g \mid < \infty$ this equation means that
the two states $\mid \v_1\rangle$ and $\mid \v_2\rangle$
are equivalent. The following question arises then: can we 
use instead of $K_i$ a more general gauge-fixing function  e.g.
a linear combination of them?
Defining now another operator by
\be
K_3 \equiv \frac{i}{2} [K_1, K_2] = \frac12 \int d^3x [A^{\parallel}(x)
E^{\parallel}(x)+ A^0(x) E^0(x) + {\cal P}(x)\y(x) - \ol{\y}(x) \ol{\cal P}(x)]
\ee
one obtains an SL(2,R) commutation algebra:
\be 
[K_1, K_3] =- i K_1, \hskip 1cm [K_2, K_3] = i K_2
\ee 
The standard form of the SL(2,R) algebra is easily recovered by
introducing the linear combinations: $X_1 = \frac{i}{2} (K_1 -K_2)$,
$X_2 = \frac{1}{2} (K_1 + K_2)$ and $X_3 = K_3$.
It should be noted here that
\be
K_1 \mid \v_2\rangle = K_2 \mid \v_1\rangle = K_3 \mid \v_1\rangle
= K_3 \mid \v_1\rangle =0
\ee
leading to the possibility of using one common gauge-fixing operator
for both $\mid \v_i\rangle$-s. This gauge-fixing operator is a
linear combination of $K_1$ and $K_2$. That is eq.(\ref{singl}) can be
equivalently rewritten in the following way:
\be
\mid s_l\rangle = e^{\a K_1 + \b K_2} \mid \v_l\rangle \hskip 1cm l=1,2   
\label{ab}
\ee
The relation between the coefficients $\a, \b$ resp. $\g_l$ in eq.(\ref{singl})
is given
as in \cite{roben} by:
\be
\g_1 = \frac{\a}{\sqrt{\a \b}} \tanh{\sqrt{\a \b}}
\ee
\be
\g_2 = \frac{\b}{\sqrt{\a \b}} \tanh{\sqrt{\a \b}}
\ee
These relations give some restrictions on the possible values
of $\a$ and $\b$, namely that both have to be non-vanishing.
If $\a \b < 0$ the $\tanh$ goes of course over to $\tan$.
What one notices now is that there exists no such pair of
finite
$\a, \b$ such that it would lead to the same $\mid s \rangle$
starting from the two different $\mid \v \rangle$.

So the answer on the question is yes, there is a large freedom
in using a linear combination of the original gauge-fixing
operators to obtain the same singlet. The next step in generalizing
the gauge-fixing operator would be to add a term containing
$K_3$. This would however only complicate the calculations
without  leading to physically new insights.

Nothing has been said yet about the {\bf time evolution} of these
various states although there seems to be a strong relationship
between evolution and gauge fixing.
The usual procedure to compute the time-evolution
of any state is to act on it
by a Hamiltonian operator defined in every space-time point:
\be
\frac{d}{dt} \mid \psi(t) \rangle 
= i \int d^3x {\cal H}(\vec{x}, t) \mid \psi(t) \rangle
\label{timedif}
\ee
The corresponding integral equation used to be written as
\be
\mid \psi(t) \rangle = e^{i \int_{t_0}^{t} dt' \int d^3x {\cal H}(\vec{x}, t')}
\mid \psi(t_0) \rangle     =  
e^{i \int_{t_0}^{t} dt' H(t')} \mid \psi(t_0) \rangle
\label{timeint}
\ee
The operator in the
exponent can be written as as
\be
\int_{t_0}^{\infty} dt' \theta (t-t') H(t')
\ee 
Then using the formal relation
\be
\theta(t-t') = \frac{1}{\partial_t} \d(t-t')
\ee one ends up with:
\be
 e^{i \frac{1}{\partial_t}  H(t)}
                                                           \label{timenew}
\ee
It is easy to see that this operator when acting on $\mid \psi(t_0) \rangle$
leads to the same
differential equation (\ref{timedif}).
Let us see how this procedure applies to our case.
The Hamiltonian obtained for the Lagrangean in equation (\ref{L}) is:
\be
H= \int d^3x [\frac{1}{2} E^\m E_\m + 
\frac{1}{2} B^i B_i + E^0 \partial_i A^i + E^i \partial_i A_0
- i \ol{\cal P} {\cal P} + i \partial_i \ol{\y} \partial^i \y]     \label{ham}
\ee
The first terms can be denoted as 
$H_1=  \int d^3x [1/2 E^\m E_\m + 1/2 B^i B_i] $, while
the rest can be written as:
\be
 H_2 = [\tilde{\r}_1 + \tilde{\r}_2, Q]
\ee
where 
\be
\tilde{\r}_1 = -i \int d^3x A^0 {\cal P}                                
\label{ro1}
\ee
\be
 \tilde{\r}_2 = \int d^3x \partial_i A^i \ol{\y}                          
\label{ro2}
\ee
and $Q$ is of course the BRST-charge given in eq.(\ref{Q}). 
One can define an operator that on the equation of motion
level is related to the time-evolution operator in (\ref{timeint}):
\be
U(t) = e^{i \int d^3x \frac{{\cal C}}{\sqrt{-\D}} {\cal H}(\vec{x}, t)} 
\label{u}
\ee
where ${\cal C}^{-1} = 4 \p \int_0^{\infty} dt (\frac{\p}{t})^{\frac32}$
is a renormalization constant.
Let us take a closer look at this operator. 

Inserting the Hamiltonian (\ref{ham}) into (\ref{u}) one obtains:
\be
U(t) =e^{i H_1'(t) + i H_2'(t)}
\ee
where
\be
H_1'(t) \equiv \int d^3x \left( \frac{1}{2 \sqrt{-\D}} E^\m E_\m
+ \frac{1}{4 \sqrt{-\D}} F^{ij} F_{ij}\right)
\ee and
\be
H_2'(t) \equiv \int d^3x \left(E^0 A^{\para} - E^{\para} A_0 
- i \frac{1}{\sqrt{-\D}} \ol{\cal P} {\cal P} 
+ i \frac{\partial_i}{\sqrt{-\D}} \ol{\y} \partial^i \y\right) \equiv K_1 + K_2
\ee
where $K_1$ and $K_2$ are the gauge-fixing operators  we used before,
i.e. the ones defined in equations (\ref{k1}) and (\ref{k2}).
We notice now that $H_1'$ and $H_2'$ are members of a closed algebra.
The other members of this algebra are defined as follows:
\be
H_3' \equiv - \frac{1}{2} [H_1', H_2'] = 
\int d^3x \frac{1}{(-\D)} E^0 E^{\para}
\ee
\be
H_4' \equiv  [H_2', H_3'] = \frac{1}{2} \int d^3x \left(\frac{1}{(-\D)} E^0 E_0
+ \frac{1}{(-\D)}  E^{\para} E_{\para}
\right)
\ee
and the only remaining non-vanishing commutator is
\be
[H_2', H_4'] = 4 H_3'
\ee
The operator
\be
H_{ph}' \equiv H_1' - \frac{1}{2} H_4' = 
\frac{1}{2} \int [\frac{1}{\sqrt{-\D}} E^i E_i -
 \frac{1}{\sqrt{-\D}} E^{\para} E_{\para}
+  \frac{1}{\sqrt{-\D}} B^i B_i]            \label{hphys}
\ee
contains only the physical variables: the orthogonal components
of the electric field and the magnetic field. Since it
commutes with all the other operators one can write
\be
U(t) = \exp{(i H_{ph}')} \exp{( iH_2' + \frac{i}{2} H_4'))}
= \exp{(i H_{ph}')} \exp{( iK_1 + iK_2 +\frac{i}{2} H_4')}  \label{udec}
\ee
We suppose now that the initial states
$\mid \v_i \rangle, i = 1,2$, defined in (\ref{fi1}) and
(\ref{fi2}) evolve in time by this operator $U(t)$.

A particularly simple way to understand what this
decomposition means is found if one goes over
to an imaginary-time formalism by making the substitution
$i t \rightarrow \t$.
  As shown in the Appendix the operator $U( \t)$ can
further be decomposed when acting on any of the $\mid \v_i \rangle$-s.
This decomposition is made possible by the fact that 
the operators in the exponent are members of an $SL(2,R)$ algebra.
For the first case one gets 
\be
\exp{  (K_1 + K_2 +\frac{1}{2} H_4')} \mid \v_1 \rangle
= \exp{ ((\tanh{1}) K_1)} 
\exp{\left(\frac{1}{2} (\tanh{1}) 
\int d^3x \frac{1}{\sqrt{-\D}} E^{\para}E_{\para}
\right)}
\mid \v_1 \rangle                    \label{h1}
\ee
and one can define a new state:
\be \mid \v_1' \rangle \equiv e^{\frac{1}{2} (\tanh{1}) 
\int \frac{1}{\sqrt{-\D}} E^{\para}E_{\para}}
\mid \v_1 \rangle\ee
which  is also a good trivial state since it is both
BRST and anti-BRST invariant. We notice that this new state is
not annihilated by all the constraints that define $\mid \v_1 \rangle$
in eq.(\ref{fi1}). Instead of the first operator ($\partial_i A^i$)
one has
\be
[\partial_i A^i(x) + i (\tanh 1) E^{\para}(x)] \mid \v_1' \rangle =0      
                                                                 \label{apara}
\ee
For $\mid \v_2 \rangle$
one gets a very similar result:
\begin{displaymath}
\exp{(K_1 + K_2 +\frac{1}{2} H_4')} \mid \v_2 \rangle =
\exp{ ((\tanh{1}) K_2)} 
\exp{\left(\frac{1}{2} (\tan{1}) \int \frac{1}{\sqrt{-\D}} E^0E_0 \right)}
\mid \v_2 \rangle \equiv 
\end{displaymath}
\be
\equiv \exp{ ((\tanh{1}) K_2)} \mid \v_2' \rangle
\ee
where $\mid \v_2' \rangle$ is again  BRST and anti-BRST invariant
and the operator that annihilates it instead of $A^0$ in (\ref{fi2}) is
\be
[A^0(x) - i {(\tanh 1)} \frac{1}{\sqrt{-\D}}E^0(x)] \mid \v_2' \rangle =0    
                                                              \label{azero}
\ee
What we found is that when acting by the imaginary-time evolution
operator on the original trivial $\mid \v_l \rangle$ states
$U(\t)$ could be decomposed as a product of three factors.
One of them $\exp(\frac{1}{2} \tanh 1 \int \frac{1}{\sqrt{-\D}} E^{\para}
E_{\para})$ resp $\exp(\frac{1}{2} \tanh 1 \int \frac{1}{\sqrt{-\D}} E^0
E_0)$ only rotated the $\mid \v_l \rangle$ states into some new trivial
states $\mid {\v'}_l \rangle$ states. The second term is just 
a gauge regulation operator $\exp ( (\tanh 1) K_l)$ and it transforms
the trivial states into physical inner-product states $\mid s_l \rangle$.
The remaining part of $U(\t)$ contains the physical Hamiltonian
and it then generates the imaginary-time evolution of 
the inner product $\mid s_l \rangle$ singlets. 
\be
\mid s(\t)\rangle =e^{ \int_{\t_0}^{\t} d{\t}' H_{phys}} 
\mid s(\t_0)\rangle\equiv
 e^{\int_{\t_0}^{\t} d{\t}' \int d^3x \frac{1}{2}[E^i E_i
-E^{\para} E^{\para} + B^i B_i]} \mid s(\t_0)\rangle
\ee
All this suggests that even in the real-time case the evolution
of the $\mid s_l \rangle$ singlets is generated by the physical
Hamiltonian (\ref{hphys}). The other arguments that relate
the time-evolution operator to the gauge-regulator are not valid 
in this case since the $\exp(i K_l) \mid \v_l \rangle$ 
states are not inner product singlets, though there exist
arguments in favour of using states like this in \cite{marnogr}. 

There exists another way too to understand these results if
one defines the inner products as propagators in the path-integral 
formalism. See e.g. \cite{annals} for the case of supersymmetric
particles, \cite{teitel} for gravity and \cite{torre} for gravity
in the Ashtekar variables. A more general picture is given in \cite{rob318}.
Here it is suggested that if the Hamiltonian in the classical
theory is of the form $H= H_{ph} + \{\r,Q\}$, the Hamilton
operator in the quantum theory should be $\hat{H} = \hat{H}_{ph}
\pm i [\hat{\r}, \hat{Q}]$. Following this suggestion
one could understand the decomposition in (\ref{udec})
the same way it was presented in the imaginary time case, namely:
the unphysical ($[\r, Q]$) part of the Hamilton operator transforms
the original non-inner product states into physical inner product
singlets and the physical part generates the time-evolution
of these singlets.
 
Another thing to be noted  is that
the original definition of the $\mid \v_l \rangle$ states
in (\ref{fi1}) and (\ref{fi2})
is by no means unique. 
These trivial states were defined by the 
complete sets of operators eliminating them.
What gives some freedom in the redefinition of these states
is the fact that  one of the operators ( the $\partial_i A^i$
for the first state and $A^0$ for the second one) 
were gauge choices and we only expected them
to complete  a BRST doublet with the constraints.
So one may always substitute the first
operators ($\partial_i A^i$ resp. $A^0$) by a linear 
combination of themselves and the constraints since these linear combinations
also satisfy the previous demand. 
This way one obtains new non-inner product states $\mid \v_l' \rangle$.
One state in both classes defined in equations (\ref{apara}) and (\ref{azero})
have very nice ``time-evolution" properties. The corresponding
part of the Hamiltonian can be substituted by the gauge-fixing
operator $K_l$.

\setcounter{equation}{0}
\section{Final Remarks}

One of the purposes of this paper was to try to find the relation
between the co-BRST operator and the gauge-fixing operators
that lead to well defined inner product states in electrodynamics.
In \cite{batrob} it was argued on general grounds that
the trivial BRST-invariant states are
not inner product states. In Section 2 we have explicitly proven this fact
for the case of free electrodynamics. We found the two 
trivial solutions $\mid \v_i \rangle, i=1,2$. The inner products
for both trivial solutions were shown to be ill-defined ($\infty 0$).

Section 3 contains a treatment of the inner product 
problem more or less in the spirit of \cite{roben}. Starting from
the two trivial BRST and anti-BRST invariant states ($\mid \v_l \rangle$)
one obtained the inner product singlets ($\mid s_l \rangle$) 
by applying on them two
gauge fixing regulators ($\exp(\g_l [\r_l, Q])$), where
$\g_l$ were free coefficients. It was shown that each singlet
could be reached from both $\mid \v_l \rangle$ states by a proper
choice of the $\g_l$ coefficients:
\be
\mid s_{\g} \rangle = \exp(\g [\r_1, Q]) \mid \v_1 \rangle
=\exp(\frac{1}{\g} [\r_2, Q]) \mid \v_2 \rangle
\ee
This means that at least in the case of Abelian models the two   
trivial solutions $\mid \v_1 \rangle$ and $\mid \v_2 \rangle$ are equivalent.
The singlets are not only inner product states but their norm is
even unity for any finite and non-vanishing $\g_l$.
As expected from \cite{batrob} the singlet states can
be given as an orthogonal set, half of them being
positive norm half of them negative norm states.
After finding these states one could construct a metric operator ($\hat{\y}$)
that would give positive norm to all the states.
This metric operator lead to the co-BRST charge as described in \cite{kalau}.
We noticed that
there is some freedom in the choice of the co-BRST operator
arising from the freedom in chosing the orthogonal
states. However once chosen the co-BRST operator
together with the BRST operator  completely
define the inner product singlet states.

In Section 4 we made
use of the fact that the gauge-fixing operators $K_1$
and $K_2$ belong to an $SL(2,R)$ algebra. In this way one could 
define a more general gauge-fixing operator as the linear
combination of the old ones: $\exp(\a [\r_1, Q] + \b [\r_2,Q])$.
It was shown that any inner product singlet $\mid s \rangle$
can be obtained this way and it was given the relation between 
$\a, \b$ resp. $\g$ leading to the same singlet.

In the last part of Section 4 we analyzed and decomposed an interesting
operator (\ref{u}) which by some very formal arguments
could be related to time evolution. 
>From this decomposition we came to the conclusion
that the time-evolution 
of the vacuum
singlets
in an imaginary-time formalism is governed by a physical Hamiltonian containing
only the orthogonal components of the electric field and
the magnetic field. 
The other terms in the Hamiltonian do not
generate imaginary-time evolution. They are the gauge-fixing
operators that transform the non-physical
$\mid \v_i \rangle$ states into the inner product $\mid s \rangle$
states. A simmilar interpretation can be made even in the
real time case if one follows the prescription
in \cite{rob318} in the construction of the Hamiltonian
operator.
The definition of the  $\mid \v_i \rangle$ 
states in (\ref{fi1}) and (\ref{fi2}) is not the 
most comfortable one from this point of view. These states were
defined by giving a complete set of operators annihilating them.
One of these operators in each case (the first ones in the definitions
(\ref{fi1}) and (\ref{fi2})) were in fact gauge choices and as such
one can always change them and obtain some new non-inner product
BRST-invariant states.
It was shown that there exists a set of  states
$\mid {\v'}_l \rangle$ defined in (\ref{apara}) and (\ref{azero})
such that the imaginary-time evolution operator
acting on them was equivalent to a gauge regulator $\exp(K_l)$.

\large{\bf Acknowlegment} 

I am very grateful to Robert Marnelius and to Igor A. Batalin
for many useful discussions.

\vskip 1cm
\newpage
\setcounter{section}{1}
\setcounter{equation}{0}
\renewcommand{\thesection}{\Alph{section}}

\noindent{\Large{\bf{Appendix}}}
\medskip

\noindent{\bf Decomposition of the Hamiltonian }

\noindent The operator acting on $\mid \v_1 \rangle$ in eq.(\ref{h1}) is
\be
e^{ i (K_1 + K_2 +\frac{1}{2} H_{41})}      \label{op}
\ee
The terms in the exponent are members of an $SL(2,R)$ algebra.
To see this one can write $H_4$  as a sum $H_4 = H_{41} + H_{42}$,
where:
\be
H_{41} = \frac{1}{2} \int \frac{1}{\sqrt{-\D}} E^0 E_0, \hskip 1 cm
H_{42} = \frac{1}{2} \int \frac{1}{\sqrt{-\D}} E^{\para} E_{\para}
\ee
Introducing the notation:
\be
a = i (K_2 + H_{41}) \hskip 1cm b = i (K_1 + H_{42}) \hskip 1cm c = -2iK_3
\ee
the operator in eq.(\ref{op}) becomes $e^{a+b}$.
The algebra these operators satisfy is:
\be
[a,b] = c, \hskip 1cm [a, c] = 2a, \hskip 1cm [b,c] = -2 b
\ee
Knowing that
\be
a \mid \v_1 \rangle = b \mid \v_2 \rangle =
c \mid \v_1 \rangle = c\mid \v_2 \rangle = 0
\ee
we can now compute how the operator is eq.(\ref{op})
acts on the different initial states. For case one we get:
\begin{displaymath}
e^{a+b} \mid \v_1 \rangle = [1+ b + \frac{1}{2!} b^2
+ \frac{1}{3!} (b^3 + 2b) + \frac{1}{4!}(b^4 + 8b^2)+
\end{displaymath}
\begin{displaymath}
+ \frac{1}{5!} (b^5 + 20 b^3 + 16 b)
+ \frac{1}{6!} (b^6 + 40 b^4 + 136 b^2) + ...] \mid \v_1 \rangle=
\end{displaymath}
\be
= e^{(\tan{1}) b}  \mid \v_1 \rangle= 
e^{i (\tan{1}) (K_1 + H_{41})}\mid \v_1 \rangle
\ee
Since $K_1$ and $H_{41}$ commute with each other one can write
\be
e^{  (K_1 + K_2 +\frac{1}{2} H_4')} \mid \v_1 \rangle =
e^{(\tanh 1) K_1} e^{ (\tanh{1}) H_{41}} \mid \v_1 \rangle
\ee
The treatment of the case 2 is trivially similar
to the first case.

\end{document}